ARTICLE

# AI based analysis of red blood cells in oscillating microchannels

Andreas Link,[a] Irene Luna Pardo,[a] Bernd Porr [a] and Thomas Franke*[a]

The flow dynamics of red blood cells in vivo in blood capillaries and in vitro in microfluidic channels is complex. Cells can obtain differnet shapes such as discoid, parachute, slipper-like shapes and various intermediate states depending on flow conditions and their viscoelastic properties. We use artificial intelligence based analysis of red blood cells (RBCs) in an oscillating mircorchannel to distinguish healthy red blood cells from red blood cells treated with formaldehyde to chemically modify their viscoelastic behavior. We used TensorFlow to train and validate a deep learning model and achived a testing accuracy of over 90%. This method is a first step to a non-invasive, label-free characterization of diseased red blood cells and will be useful for diagnostic purposes in haematology labs. This method provides quantitative data on the number of affected cells based on single cell classification.

## Introduction

Blood test still play a major role in diagnostics since samples are easy to obtain and are routinely used for analysis in clinical and general practice setting.[1] Red blood cells (erythrocytes) form by far the major component of whole blood (about 50 %) and have been used as biological indicators for several diseases such as sickle cell anaemia, spherocytosis, beta-thalassemia, malaria,[2,3] hypercholesterolemia[4,5] and many others. In many of these diseases the shape of the erythrocyte is altered, so simple optical microscopical examination and cell counting allows quantitative analysis in stasis e.g., in a blood smear. The adopted shape depends on mechanical properties of the cell and the external strain. In hydrodynamic flow, there have been various shapes been reported in vivo as well as in vitro. For healthy cells these encompass discocytes, parachutes and slipper-like shapes and echinocytes, spherocytes, etc for diseased cells. However, not all cells of a sample are in the same shape condition, due to variances in mechanical properties within the cell population. Even within one blood sample there exists a variance of mechanical properties of the erythrocyte since for example cells alter their properties and size during their ~120 days lifetime of circulation in the organism.[6] Red blood cells (RBCs) in microflow conditions have been analysed both in experiments[7] and theory.[8] Most of the studies are in capillary flow, yet some research was done on more complex geometries to reveal the shape relaxation dynamics.[9,10]
Based on viscoelastic models the experimentally observed cells could be reproduced and variation of mechanical parameters (viscoelastic moduli) were determined has also demonstrated that transitions in between these cell shapes can occur.

However, in experimental settings it is still challenging to distinguish between populations of red blood cells from different samples, such as diseased cells and healthy cells.
In AI based red blood cell classification most of the prior work[11–13] distinguished between populations of cells with the associated problems mentioned above and in addition poses technical problems when employing segmentation techniques to isolate individual blood cells which are error prone and computationally intensive. Because there is a strong interest in the viscoelastic properties of the red blood cells often only their shape has been analysed but not the healthiness of the cells.[14,15] The use of a channel to probe the mechanical properties of the red blood cells in conjunction with AI to detect a disease has, to our knowledge, only attempted once.[16]

Here, we demonstrate an AI based image analysis using TensorFlow that can decide between native, untreated red blood cells and red blood cells with chemically altered mechanical properties. Using a microchannel with oscillating width we transiently deform cells to capture both, the elastic, and the viscous properties. Our device presents a method that can distinguish a population of cells which are not obvious by simple inspection by eye. It is ready to be used for disease diagnosis and analysis of the severity of a disease by providing quantitative results on the frequency of affected cells.

## Experimentals and Results

### Microfluidic procedures / setup

We probe red blood cells in flow with periodically oscillating flow velocities in a zigzag-shaped microchannel. The channels are fabricated using standard soft-lithography and mounted on an inverted fluorescence microscope. Videos of cells flowing through the channel are recorded with a high-speed camera (Photron, UX50). Red blood cell samples are prepared from whole blood by centrifugation and incubation steps with chemicals for modification of their mechanical properties (see material and methods section). We use a diluted red blood cell suspension in PBS buffer with a haematocrit ($Ht$) of $Ht = 0.5\ \%$ and inject it into the inlet of the microfluidics device using a pressure driven system, controlling the pressure drop between inlet and outlet. Cells then enter a zigzag-shaped region that is in the field of view of a 40 x objective as shown in Fig. 1. To avoid large sections of videos without cells and to reduce memory consumption at the low chosen, we trigger the recording of videos by a hardware trigger. Therefore, we use a laser to excite (488 nm) a fluorescent marker in the red blood cell and detect the emitted (525 nm) fluorescence signal with a photomultiplier tube (PMT). In this way we record cell sequences of 60 frames per cell. We record 250 sequences and then transfer the batch of data to the computer hard drive for post-processing.

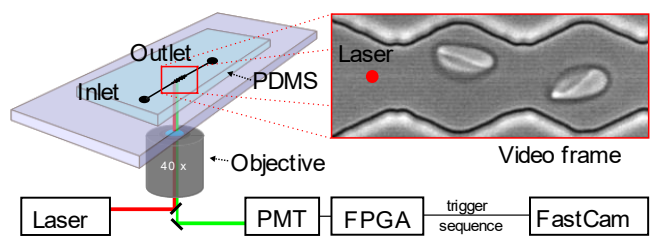

Figure 1 Schematic of the microfluidic setup mounted on an inverted fluorescence microscope: RBCs entering the device and flow into the region of oscillating width and adopt their shape. Videos are taken with a fast camera and recording is triggered by cell passing a photoelectric barrier and rapid analysis in a photomultiplier tube.

**Video processing**

Three batches of 248 sequences each were recorded for healthy cells which results in a total of 744 video sequences of native red blood cells (the first two sequences where removed). Two batches of 248 sequences each were recorded for chemically modified cells which resulted in a total of 496 video sequences. Fig. 2 shows how the training data has been extracted from the raw video sequences. Fig. 2A shows a snapshot of one clip where the cells are flowing from the left to the right. Here, the cell has already progressed to the 2nd narrow section of the channel. Since we are interested how the zigzag-channel impacts on the cell's behaviour and shape, we have defined a "soft trigger" where a snapshot for training is used. One can think of the soft trigger as a finishing line in a race. As an example, we have shown one soft trigger when the cell in the red square has just arrived at a narrow section and has touched the soft trigger. A 2nd soft trigger has also been established (not shown) when the cell has arrived at a wide section and thus, we have two soft triggers: one for taking a snapshot in the narrow section and one for the section. Once the cell has been detected a snapshot is taken of the size of the crop window indicated with "crop" in Fig. 2A. As a next step the background was obtained (Fig. 2B) and subtracted (Fig. 2C) to prevent TensorFlow from learning features of the background instead of that of the red blood cells. Since the cell flows through the image from left to right, we can take the background for the left half from the final frames of a clip and the background from the right half from the first frames of the clip. Splicing these two halves together gives the background which can then be used in Fig. 2C to obtain a background-free cell image.

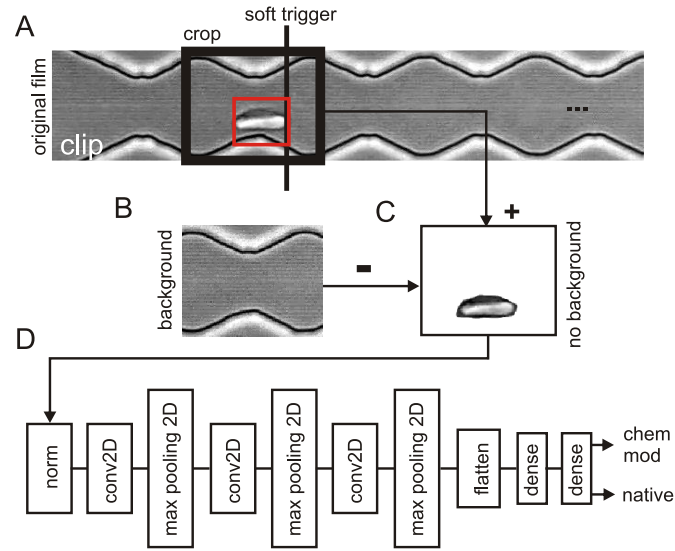

Figure 2 Data processing pipeline. (A) Single frame from a video clip with a red blood cell in the red boundary box touches the "soft trigger" (black line) which then leads to the extraction of the image in the area "crop" (black box). (B) Background of the cropped area from (A) by taking a video frame from the video clip without a red blood cell. (C) Background subtraction of (B) subtracted from (C) and taken the absolute value. (D) TensorFlow layers: "norm": RGB value normalisation to 0.01, "conv2D": standard 2D convolutional layer, "max pooling": standard 2D max pooling, "dense": standard dense layer. The final layer has two outputs: one for the detection probability of a native red blood cell and one for the chemically modified one.

**Data analysis and decision making with TensorFlow**

The video processing pipeline described above creates three batches of 248 images (one image per sequence) of native red blood cells and two batches of 248 images of chemically modified red blood cells for training. For training the model every image is labelled of either being healthy or diseased. This results in $5x248$ images/label-pairs. These pairs are then shuffled randomly for training and then fed into TensorFlow (Fig. 2D). This is a standard network topology as suggested by TensorFlow/Keras to do image classification. The network has two outputs, one gives the probability of being a native cell and one the probability of being a chemically modified one. Training was performed with 5 epochs and the accuracy evaluated with two separate batches of clips not being part of the training: 492 images of healthy red blood cells and 947 images of diseased cells. These two subpopulations where then sent into TensorFlow for training (80 %) and validation (20 %) which is described next.

**Comparison of frame location in the microfluidic channel**

To investigate the impact of the position of the cell in the microchannel we created two datasets of red blood cell pictures and fed them into TensorFlow – one with the red blood cells detected in the widening section of the microchannel and another one with the red blood cells in the narrow section as shown in Fig. 3. The goal of training is to have one output of the classifier close to one and its other output close to zero so that either the "native" output becomes one or the "chemically modified" becomes one but never both. TensorFlow training and validation on the dataset containing images from the narrow part of the channel is shown in Fig. 3A: on the left the training and validation accuracy is shown and on the right the training and validation loss is shown. Both training and validation accuracy reach 99 % after 5 epochs and stabilise there. The same applies for the training and validation loss which reach their minimum after 5 epochs. Training using images from the wide section (Fig. 3B) of the microchannel result in remarkably similar behaviour: both the training accuracy and the validation accuracy converge to over 99% after 5 epochs. Also, the training loss and validation loss reach their minima in a similar way as the training in the narrow section.

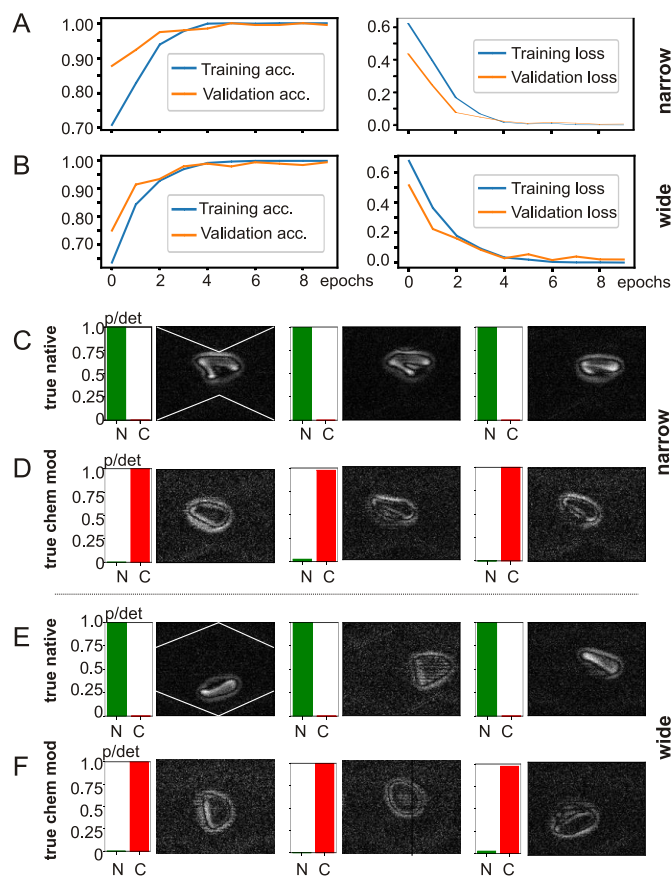

Figure 1 TensorFlow training behaviour and results. (A) Training and validation loss for the narrow part of the microchannel and for the wide part (B) of the microchannel. Classification results for the narrow section of the microchannel for (C) native red blood cells and (D) chemically modified red blood cells. Classification results for the wide section of the microchannel for (E) native red blood cells and (F) chemically modified red blood cells. "N": native, "C": chemically modified, "p/det": detection probability.

After training the TensorFlow model was presented with test data which the network has not yet seen: 492 clips of native red blood cells and 111 clips of chemically modified cells. The testing accuracy at the narrow section was 98 % and for the wide one 97 %. Fig. 3C/D show the testing results for the dataset generated at the narrow section of the channel and Fig. 3E/F at the wide section of the channel. Every row shows three examples of red blood cells and their detection probability against the ground truth: true native or true chemically modified. Remember that training forces the network to clearly output a one for its detection category. For example, in Fig. 3C the ground truth is "native" and the network outputs a one for native, indicated as "N" under the green bar. Then in Fig. 3D the cells were chemically modified, and the network classified them all with very high confidence as chemically modified ("C"). Fig. 3E/F now shows the same as above but for the wide section of the channel. Again, true native ones are detected as native without any doubt and reflects the 97 % of the testing accuracy. The truly chemically modified ones were also detected with ease. Having just two categories being native or chemically modified allows a simple comparison which is robust with one category being almost zero and the other almost one.

## Material and Methods

### Microfluidic device preparation

Microfluidic channels are fabricated using soft lithography.[17,18] Briefly, the channel structure is designed in CAD program and transferred to a chromium mask (ML&C GmbH). The zigzag-shaped channel has a periodicity of 20 μm, and an amplitude of 20 μm. The narrow part is 10 μm wide (for details see sketch in the SI). To structure of the mask is transferred to a silicon wafer coated with 10 μm SU8-3010 photoresist (Microchem, SU8 3000 series) using a mask aligner (MA6, Süss MicroTec). After developing with (Microposit™EC Solvent, Shipley), the structured SU8 acts as a template to fabricate PDMS moulds (PDMS, Sylgard™ 184 Silicone Elastomer Kit). The fluid PDMS is poured onto the template and cured for 4h in an oven at 75°C. The ratio of elastomer base to curing agent is 10: 1

We punch holed into the cured PDMS mould to connect inlet and outlet to the tubing. The PDMS is covalently bonded onto a microscope slide using oxygen plasma.

### RBC preparation

Whole blood was obtained from healthy donors (Research Donors, Cambridge Bioscience) and washed three times in a phosphate-buffered saline solution (PBS 1×, pH 7.4, 330 mOsm/L, Gibco Life Technologies). After each washing step, the sample was centrifuged for 5 minutes at 2500 RPM (mini spin plus centrifuge, Eppendorf), and the white buffy coat and supernatant were removed.

For the native red blood cell sample, we used 2 μL of concentrated RBC pellet from the bottom of the reaction tube and incubated in 0.5 mL green-fluorescent calcein-AM solution (5 μM, Invitrogen) for 30 minutes at 37°C, followed by three washes with PBS.

For the chemically modified RBC experiments, we first incubate the red blood cells in formaldehyde solution before we stain the cells. Briefly, 10 μL of the pellet were incubated in 0.37 % formaldehyde solution (Sigma-Aldrich) for 10 min at room temperature. The cell suspension was then washed three times before stained in calcein-AM as described above.

For the experiment 5 μL of the native or chemically modified cell suspension were mixed with 995 μL of PBS solution and used as the stock solution with a haematocrit of 0.5 %. The experiments were completed within a day of blood collection.

To avoid cell sedimentation during the experiment, the cells were suspended in a density-matched solution using a density gradient solution (OptiPrep Density Gradient Medium, Sigma Life Science). Optiprep is a sterile non-ionic solution of 60 % (w/v) iodixanol in water. Additionally, bovine serum albumin (BSA, Ameresco) was used to prevent cell adhesion to each other and the microchannel walls. Therefore, 473 μL of OptiPrep solution was mixed with 1527 μL of PBS/BSA (14 mg/mL) and suspended in 20 μL of concentrated fluorescent-labelled RBC solution. The final density of this solution was $\rho = 1.080$ g/mL, and the estimated haematocrit ($Ht$) of this solution was 0.5 %.

**Python code**

The red blood cells videos consist of multiple clips and every clip has a specific number of frames. We identified two "finishing lines" at different positions in the videos where we wanted to capture images of cells at the wide and narrow sections. The finishing line was different for wide and narrow section at each of the videos and it was inputted as one of the values.

To detect the cells at these finishing lines, we used OpenCV background subtractor, a Gaussian Mixture-based Background/Foreground segmentation, with a history of 100 and varThreshold of 20. We exclusively applied this only to the specific detection area which was determined by the finishing line parameter we had specified. The detection area was defined by a y range of (10,110) and an x range from the finishing line to the finishing line plus 250. Then, we used OpenCV function "findcontours" to create contours around the moving cells in that detection area. If these contours were bigger than an area of 20 units squared, we recorded the frame number and extracted an image.

The resulting images were cropped to a size of 120 x 150 pixels, with a cropping range of 0 to 120 for height and from the finishing line minus 80 to the finishing line plus 70 for the width. Therefore, images were of a shape of (120,150,3).

For our neural network, we used Keras sequential model with 10 layers in the following order, including Rescaling, 2D convolutional layer with 16 as the number of output filters, Maxpooling2D layer, 2D convolutional layer with 32 as the number of output filters, Maxpooling2D layer, 2D convolutional layer with 64 as the number of output filters, Maxpooling2D layer, Flatten layer, Dense layer with unit value 128 and, lastly, Dense layer with unit value of 2 which are both labels healthy and unhealthy. The 2D convolutional layers have a stride parameter of (1,1) and a kernel size of (3,3). For the training we used a number of epochs of 10 and a batch size of 32.

## Discussion and Conclusion

In this study we have used periodically oscillating zigzag-shaped microchannels to probe RBCs. Compared to many other studies the spatially alternate design enables the interrogation of viscous cell properties. In contrast, RBCs in cylindrical capillary flow adopt a constant shape and neither dynamic, viscous properties of the RBC cell membrane nor the cytosol enter because the cell moves as a fixed object in flow. Here, we probed cells at two separate locations in microflow, at the widening and the constriction. In between these extrema, the RBCs adopt an intermediate shape. The transition between these shapes in the wide and narrow section is controlled by the relaxation time of the RBC as given by the ratio of viscous to elastic parameters. Here, we tested native and chemically modified RBCs using formalin at 0.3% volume ratio (formaldehyde in water solution). Formalin is a fixation agent and non-specifically cross-links proteins in the RBC thereby modifying the viscoelastic, mechanical properties of the cell. Similar to glutaraldehyde (GA), the aldehyde group binds and can interlink amino acids of proteins and thereby mechanically stabilize the RBC membrane. In GA mediated mechanical modification, the GA can exist as a monomeric or polymeric form, both of which bind to amino acids. Therefore, GA can link proteins over a variable distance depending on the length of the polymer. Even though formalin can also form formaldehyde polymers, the mechanically reactive form with an aldehyde group is the monomer, i.e. methylene hydrate. Only as monomer it can bind to proteins to form a methylene cross-link. Therefore, the potential for cross-linking with GA is much larger, firstly because of its two aldehyde binding sites and secondly due to the variable lengths (cite Abay2019). Hence, the effect of formalin is milder and slower than the one of GA, facilitating chemical control of mechanical RBC properties. However, due to the small size formalin has been reported to have a much higher permeation rate compared to GA (cite Kiernan). It can potentially enter the RBC and link proteins in the cytosol, such as hemoglobin.[19]

In AFM studies an increase in elastic modulus depending on GA concentration has been reported as well as a reduction in deformability (elongation index). (cite Abay2019) AFM measurements using formalin at a concentration of 5% have revealed a 10-fold increase in Young modulus as compared to untreated cells.[20] The difference in stiffness leads to less deformation and a lower elongation index as obtained from a shape analysis.[21]

Stiffening of RBC membrane has also been achieved by incubation in diamide.[22–24] However, diamide has been reported to only rigidify the membrane stiffness and fluidity. It provides disulphide bridges between specific thiol-group containing amino-acids (cystine, methionine) and has a minor effect on the cytosol. Its effect has been studied in microflow analysing the shape changes by using the cell width to length ratio.[25]

In our experiments, the changes in viscoelastic properties due to chemical treatment with formalin 0.3% are expected to affect the morphology of the RBC in flow. However, we could not detect any apparent shape changes in our micrographs by eye or simple analysis such as analysing the deformation index or projected area.

Instead of analysing the RBC shape and contour as we have done in previous studies,[26,27] here we have taken an AI based approach.

Using a TensorFlow based machine learning we categorize native and chemically treated cells with high precision. This goes beyond current literature that probe red blood cells in capillary flow.[16]

Central to red blood cell diagnostics is the analysis of their deformability and, thus, their ability to form different shapes.[16,28] For example, Kihm et al. divided the cells in two groups called "Slippers" and "Croissants",[14] implemented a CNN (Convolutional Neural Network) to train 4000 images and classify the RBCs uniquely based on their shape characteristics. Considering the increasing demand for advancements and the potential for significant impact and popularity in this field, Recktenwald et al.[29] and its follow-up study[30] adopted the approach, proposed by Kihm et al. and Alkrimi et al.,[14,15] to benchmark different AI techniques classifying RBCs and similarly to Kihm et al.,[14] the classification was based fully on morphology. Lee at al.[31] uses not just shape but also texture features to classify normal and abnormal RBCs and similarly to Das et al.,[12] they classify cells in more than two categories. However, overall, these studies employ an intermediate step by first or exclusively focussing on morphology and then feeding the pre-processed data into the final classifier for diagnosis. However, this assumes that one knows which deformation, shape or texture relates to a diseased cell and which one to a healthy one. In contrast, we do pixel-to-disease-classification (i.e., end to end) where the deep net learns the distinguishing features by itself without 1st hand-crafting features and only then feeding them into a classifier.

Often raw microscopic images taken contain a large amount of red blood cells which requires a segmentation process to extract individual images.[16,31] Due to the segmentation processes and individual image extraction, the quality of the image is often seen reduced and as a consequence degrades the classification accuracy. Other approaches to work with images of many red blood cells are the Circular Hough transform[32] or region-based segmentation (ORBS).[13] Das et al.[12] also studies segmentation of RBCs in image classification and similarly to the other studies[32] and the study of Shemona et al.,[13] the images analysed also contain many RBCs. In contrast to these studies our work directly captures single red blood cells passing through the microchannel and is not affected by image degradation, low resolution after cropping or wrong segmentation.

AI can be used to classify RBCs into more than two categories,[12,31] for example Malaria, Thalassaemia, other abnormal and normal.[11] In our study we have used two categories as a proof of concept but the TensorFlow classifier can take any number of classes where the only limit is the available computing power and GPU memory. Future work will use more than two categories.

Confounding factors are a significant challenge in any AI based learning algorithm which is at risk of learning just trivial and superficial features from the training data. For instance, Zech et al. explains that CNNs may not effectively identify disease-specific finings but rather exploit confounding information.[33] This crucial issue is only addressed by Rizzuto et al. which tries to eliminate any confounding structures cropping the video to extract only a limited area of interest.[16] No study has explored the effect of confounding factors such as focal plane on image classification of RBCs. In this study we discovered that AI would use the focal plane to distinguish between native and chemically modified RBCs when using one focal plane for the native ones and one for the diseased ones. To overcome this problem, we mixed images captured at different focal planes and, in addition, removed the background from the images.

# References

1  M. Gary, G. Knight and A. D. Blann, *Haematology*, Oxford University Press, Oxford, Third ed., 2021.
2  T. M. Geislinger, S. Chan, K. Moll, A. Wixforth, M. Wahlgren and T. Franke, *Malar. J.*, 2014, **13**, 375.
3  H. A. Cranston, C. W. Boylan, G. L. Carroll, S. P. Sutera, J. R. Williamson, I. Y. Gluzman and D. J. Krogstad, *Science (80-. ).*, 1984, **223**, 400–403.
4  M. Kohno, K. Murakawa, K. Yasunari, K. Yokokawa, T. Horio, H. Kano, M. Minami and J. Yoshikawa, *Metabolism*, 1997, **46**, 287–291.
5  J. Radosinska and N. Vrbjar, *Physiol. Res.*, 2016, **65**, S43-54.
6  M. J. Simmonds, H. J. Meiselman and O. K. Baskurt, *J. Geriatr. Cardiol.*, 2013, 10, 291–301.
7  G. Tomaiuolo, M. Simeone, V. Martinelli, B. Rotoli and S. Guido, *Soft Matter*, 2009, 5, 3736–3740.
8  H. Noguchi and G. Gompper, *Proc. Natl. Acad. Sci.*, 2005, **102**, 14159–14164.
9  S. Braunmüller, L. Schmid and T. Franke, *J. Phys. Condens. Matter*, 2011, **23**, 184116.
10 S. Braunmüller, L. Schmid, E. Sackmann and T. Franke, *Soft Matter*, 2012, **8**, 11240–11248.
11 Y. Hirimutugoda and G. Wijayarathna, *Sri Lanka J. Bio-Medical Informatics*, 2010, **1**, 35.
12 D. K. Das, C. Chakraborty, B. Mitra, A. K. Maiti and A. K. Ray, *J. Microsc.*, 2013, **249**, 136–149.
13 J. S. Shemona and A. K. Chellappan, *IET Image Process.*, 2020, **14**, 1726–1732.
14 A. Kihm, L. Kaestner, C. Wagner and S. Quint, *PLOS Comput. Biol.*, 2018, **14**, e1006278.
15 J. A. Alkrimi, S. A. Tome and L. E. George, *Eur. J. Eng. Res. Sci.*, 2019, **4**, 17–22.
16 V. Rizzuto, A. Mencattini, B. Álvarez-González, D. Di Giuseppe, E. Martinelli, D. Beneitez-Pastor, M. del M. Mañú-Pereira, M. J. Lopez-Martinez and J. Samitier, *Sci. Rep.*, 2021, **11**, 13553.
17 Y. Xia and G. M. Whitesides, *Annu. Rev. Mater. Sci.*, 1998, **28**, 153–184.
18 D. Qin, Y. Xia and G. M. Whitesides, *Nat. Protoc.*, 2010, **5**, 491–502.

19  J. A. Kiernan, *Micros. Today*, 2000, **8**, 8–13.
20  T. G. Kuznetsova, M. N. Starodubtseva, N. I. Yegorenkov, S. A. Chizhik and R. I. Zhdanov, *Micron*, 2007, **38**, 824–833.
21  T. Go, H. Byeon and S. J. Lee, *Sci. Rep.*, 2017, **7**, 41162.
22  N. Mohandas, M. R. Clark, M. S. Jacobs and S. B. Shohet, *J. Clin. Invest.*, 1980, **66**, 563–573.
23  A. M. Forsyth, J. Wan, W. D. Ristenpart and H. A. Stone, *Microvasc. Res.*, 2010, **80**, 37–43.
24  I. Safeukui, P. A. Buffet, G. Deplaine, S. Perrot, V. Brousse, A. Sauvanet, B. Aussilhou, S. Dokmak, A. Couvelard, D. Cazals-Hatem, O. Mercereau-Puijalon, G. Milon, P. H. David and N. Mohandas, *Blood Adv.*, 2018, **2**, 2581–2587.
25  M. Faivre, C. Renoux, A. Bessaa, L. Da Costa, P. Joly, A. Gauthier and P. Connes, *Front. Physiol.*, 2020, **11**, 1–10.
26  H. Noguchi, G. Gompper, L. Schmid, A. Wixforth and T. Franke, *EPL (Europhysics Lett.*, 2010, **89**, 28002.
27  S. Braunmüller, L. Schmid and T. Franke, *J. Phys. Condens. Matter*, 2011, **23**, 184116.
28  E. S. Lamoureux, E. Islamzada, M. V. J. Wiens, K. Matthews, S. P. Duffy and H. Ma, *Lab Chip*, 2022, **22**, 26–39.
29  S. M. Recktenwald, M. G. M. Lopes, S. Peter, S. Hof, G. Simionato, K. Peikert, A. Hermann, A. Danek, K. van Bentum, H. Eichler, C. Wagner, S. Quint and L. Kaestner, *Front. Physiol.*, , DOI:10.3389/fphys.2022.884690.
30  S. M. Recktenwald, G. Simionato, M. G. M. Lopes, F. Gamboni, M. Dzieciatkowska, P. Meybohm, K. Zacharowski, A. von Knethen, C. Wagner, L. Kaestner, A. D'Alessandro and S. Quint, *Elife*, , DOI:10.7554/eLife.81316.
31  H. Lee and Y.-P. P. Chen, *Pattern Recognit. Lett.*, 2014, **49**, 155–161.
32  H. A. Elsalamony, *Micron*, 2016, **83**, 32–41.
33  J. R. Zech, M. A. Badgeley, M. Liu, A. B. Costa, J. J. Titano and E. K. Oermann, *PLoS Med.*, 2018, **15**, e1002683.